\documentclass{ws-procs975x65}

\begin{document}

\begin{flushright}
IPMU 10-0058
\end{flushright}

\bigskip

\title{POSSIBLE SOLUTION OF DARK MATTER, THE SOLUTION
OF DARK ENERGY AND GELL-MANN AS GREAT THEORETICIAN}

\author{Paul Howard Frampton$^*$}

\address{Department of Physics and Astronomy,
University of North Carolina, Chapel Hill, USA {\it and}
Institute for the Mathematics and Physics of the Universe, University of Tokyo,
Japan.\\
$^*$E-mail: frampton@physics.unc.edu {\it and} 
paul.frampton@ipmu.jp}

\bigskip

\begin{abstract}
This talk discusses the formation of primordial intermediate-mass
black holes, in a double-inflationary theory, of sufficient abundance
possibly to provide all of the cosmological dark matter. There follows my,
hopefully convincing, explanation of the dark energy
problem, based on the observation that the visible universe
is well approximated by a black hole.
Finally, I discuss
that Gell-Mann is among the five greatest
theoreticians of the twentieth century.
\end{abstract}

{\tt This work is dedicated to Murray Gell-Mann for his 80th birthday.}


\bodymatter

\section{Outline of talk}
\noindent
It is an honor to talk at a festschrift for Murray
Gell-Mann, who dominated research
in particle phenomenology for at least twenty years.

\bigskip

\noindent
At the beginning of my talk, I shall discuss a recent 
paper \cite{FKTY} on the production
of primordial intermediate-mass black holes of mass $M_{BH}=M_{\odot}^p$
with $-8 \leq p \leq +5$, providing a sufficient abundance, that the
primordial IMBHs can possibly act as all the cosmological dark matter.
\bigskip

\noindent 
I then discuss my solution\cite{Frampton} for the difficult dark energy problem
which was first identified from observations of
supernovae, twelve years ago. Although I knew all the correct
theoretical ingredients back then, the solution hit me only
on February 6, 2010. Because this was
an overwhelming human experience, I self-indulgently
discuss it.

\bigskip

\noindent
Finally, I discuss why Gell-Mann, who must himself have experienced
a similar personally fulfilling moment, for the
$\Omega^{-}$ particle\cite{MGMomega,EXPTomega}, is to be correctly,
regarded as among the five greatest theoreticians, of the
twentieth century. 

\bigskip

\section{Possible Solution for Dark Matter}

\noindent
If the dark matter (DM) is made of a weakly interacting massive particle (WIMP), we
may be able to observe collider, direct and indirect DM signatures;
the DM particles may be produced at LHC, and the next-generation
direct search experiments will probe a significant portion of
parameter space predicted by various theoretical DM models.  In spite
of thorough DM searches using widely different techniques, the results
are negative so far. If no DM signature is found in the
future experiments, it may suggest that the basic assumption that the
DM is made of unknown particles is simply wrong.

\bigskip

\noindent
There actually {\it is} a DM candidate in
the framework of SM, namely, a primordial black hole (PBH). 
In the early Universe PBHs can form when
the density perturbation becomes large, and it has been known that a
PBH of mass greater than $10^{15}$\,g survives
evaporation, and therefore contributes to the DM
density.

\bigskip

\noindent
In consideration of the entropy of the universe it was pointed out in 
Ref.~\cite{Frampton:2009nx}
that if all DM were in the form of $10^5 M_{\odot}$ black holes it
would contribute a thousand times more entropy than the supermassive
black holes at galactic centers and hence be a statistically favored
configuration. Here we consider primordial black holes (PBHs) with masses from 
$10^5 M_{\odot}$ to $10^{-8} M_{\odot}$ and, subject to observational 
constraints, any of these masses can comprise all DM although 
the entropy argument favors the heaviest $10^5 M_{\odot}$ mass.

\bigskip

\noindent
There are several ways to realize large density fluctuations leading
to PBH formation.  One possibility is the production of PBHs from
density fluctuations generated during inflation.  Since the blue
spectrum with a spectral index $n_s > 1$ is disfavored by the WMAP
data, a single inflation may not be able to produce large
density fluctuations at small scales unless some dynamics is
introduced during inflation.  On the other hand, the density
fluctuations can be easily enhanced at small scales in a double
inflation model.

\bigskip

\noindent
In Ref. [\cite{FKTY}], we discuss a double inflation model that consists of a
smooth-hybrid inflation and a new
inflation. In this set-up PBHs with a narrow
mass distribution are formed as a result of an explosive particle
production between the two inflations. We show that
the PBH mass can take a wide range of values from $10^{-8} M_\odot$ up
to $10^5 M_\odot$.  Also, the resultant PBH mass has a correlation
with running of spectral index. 
We numerically calculated the correlation, which can be
tested by future observations.

\bigskip

\noindent
The black hole mass, and the formation epoch, are related to each other,
due to the causality.  In the early Universe, the mass contained in
the Hubble horizon sets an upper bound on the PBH mass formed at that
time.  Assuming that the whole mass in the horizon is absorbed into
one black hole, we obtain

\begin{eqnarray}
M_{\rm BH} &=&\frac{4 \pi \sqrt{3} M_P^3}{\sqrt{\rho_f}}
 \simeq 0.05\, M_\odot \frac{g_*}{100}{-\frac{1}{2}} \frac{T_f}{{\rm GeV}}{-2},\nonumber \\
 & \simeq &1.4 \times 10^{13}\, M_\odot   \frac{g_*}{100}{-\frac{1}{6}} \frac{k_f}{{\rm Mpc}^{-1}}{-2},
 \label{mpbh}
\end{eqnarray}

\noindent
where $M_{\rm BH}$ is the black hole mass, $M_P \simeq 2.4 \times
10^{18}$\,GeV is the reduced Planck mass, $M_\odot \simeq 2 \times
10^{33}$\,g is the solar mass, $g_*$ counts the light degrees of
freedom in thermal equilibrium, $\rho_f$, $T_f$ and $k_f$ are the
energy density, the plasma temperature and the comoving wavenumber
corresponding to the Hubble horizon at the formation, respectively.
The radiation domination was assumed in the second equality.

\bigskip

\noindent
As is well known, any black holes have a temperature
inversely proportional to its mass and evaporates in a finite time
$\tau_{\rm BH}$,

\begin{equation}
\tau_{\rm BH} \;\simeq\;10^{64} \frac{M_{\rm BH}}{M_\odot}{3} {\rm yr}.
\end{equation}

\noindent
Thus the black holes with mass less than $10^{15}$ g must have
evaporated by now.  PBHs which remain as (a part of) DM must therefore
be created at a temperature below $10^{9}$ GeV. In the following we
assume that PBHs account for all DM in our Universe.

\bigskip

\noindent
The cosmological effects of PBHs have been extensively studied so
far. While PBHs with masses below $10^{15}$ g are significantly
constrained, it is very difficult to detect PBHs heavier than
$10^{15}$ g because of negligible amount of the radiation.
The MACHO and EROS
collaborations monitored millions of stars in the Magellanic Clouds to
search for microlensing events caused by MAssive Compact Objects
(MACHOs) passing near the line of sight.  The MACHO
collaboration excluded the objects in the mass
range $0.3 M_\odot$ to $30 M_\odot$, and the latest result of the
EROS-1 and EROS-2 excluded the mass range $0.6
\times 10^{-7} M_\odot < M < 15 M_\odot$, as the bulk component of the
galactic DM. On the other hand, if we assume that the PBH formation
occurs before the big bang nucleosynthesis (BBN) epoch, the PBH mass
should be lighter than $10^5 M_{\odot}$.
Therefore we consider PBHs with masses (i) $M_{\rm BH} < 10^{-7}
M_\odot$ and (ii) $30 M_{\odot} <M_{\rm BH} < 10^5 M_\odot$.

\bigskip

\noindent
The above observational constraints provide us with information on the
PBH formation.  If PBHs are produced at different times, the mass
function tends to be broad, thereby making it difficult to be
consistent with observations. In order to realize the PBH mass
function with a sharp peak, most of the PBHs should be produced at the
same time.  Thus the production mechanism must involve such a dynamics
that only the density fluctuation of a certain wavelength rapidly
grows.

\bigskip

\noindent
What kind of dynamics can create PBHs?  First of all, density
perturbation must become large for PBHs to be formed.  There are
several ways to realize large density fluctuations leading to the PBH
formation.  One possibility
is the production of PBHs from density fluctuations generated during
inflation. In the standard picture of inflation, the inflation driven
by a slow-rolling scalar field lasts for more than about $60$
e-foldings to solve theoretical problems of the big bang
cosmology. Then no dynamics for producing a sharp peak in the density
perturbation is expected.  However, there is no
a priori reason to believe that our Universe experienced only one
inflationary expansion. Indeed, the cosmological gravitino or modulus
problem can be relaxed if the energy scale of the last inflation is
rather low, and it is then quite likely that there was another
inflation before the last one. If the multiple inflation is a common
phenomenon, we expect that explosive particle production between the
successive inflation periods may produce a sharp peak in the density
perturbation at the desired scales, which leads to the PBH formation
at a later time. In the next section, we show that this is actually
feasible using a concrete double inflation model.

\bigskip

\noindent
We provide a double inflation model, producing PBHs
with a sharp mass function, as an existence proof. The 
first inflation is realized by smooth hybrid
inflation.  The smooth hybrid
inflation model is built in framework of supergravity and the
superpotential and K\"ahler potential are given by

\begin{eqnarray}
  W_{H} & = & S\left(\mu^2 + \frac{(\bar{\Psi}\Psi)^m}{M^{2(m-1)}}\right)
  ~~~~~(m= 2,3,\ldots),\\
  K_H & = & |S|^2 + |\Psi |^2 + |\bar{\Psi}|^2,
\end{eqnarray}

\noindent
where $S$ is the inflaton superfield, $\Psi$ and $\bar{\Psi}$ are
waterfall superfields, $\mu$ is the inflation scale and $M$ is the
cut-off scale which controls the nonrenormalizable term.  From the
above superpotential and K\"ahler potential together with phase
redefinition and the D-flat condition, we obtain the scalar potential
as

\begin{equation}
   V_H(\sigma, \psi) \;\simeq\;  \left(1+\frac{\sigma^4}{8}+ \frac{\psi^2}{2}\right)
   \left(-\mu^2 + \frac{\psi^{4}}{4M^{2}}\right)^2
   + \frac{\sigma^2\psi^6}{16M^4},
   	\label{eq:pot_hybrid}
\end{equation}

\noindent
where $\sigma \equiv \sqrt{2}Re S$ and $\psi\equiv 2Re \Psi = 2Re
\bar{\Psi}$.  Here and in what follows we use the Planck unit $M_P=1$
and take $m=2$ for simplicity.  Although the scalar potential
(\ref{eq:pot_hybrid}) is derived in the framework of supergravity, one
may start with (\ref{eq:pot_hybrid}) without assuming supersymmetry.
The potential (\ref{eq:pot_hybrid}) has a true vacuum at $\sigma = 0$
and $\psi = 2\sqrt{\mu M}$. For $\sigma \gtrsim \sqrt{\mu M}/2$,
however, the potential for $\psi$ has a $\sigma$-dependent minimum at

\begin{equation}
   \psi_{\rm min} \;\simeq\; \frac{2}{\sqrt{3}}\frac{\mu M}{\sigma}.
   	\label{eq:path_hybrid}
\end{equation}

\noindent
Note that $\psi$ quickly settles down at the minimum during inflation
since its mass is larger than the Hubble parameter.  Then we can
integrated out $\psi$ and obtain the effective potential for $\sigma$
as

\begin{equation}
   V(\sigma) \;=\; \mu^4 \left( 1 + \frac{\sigma^4}{8}
   - \frac{2}{27} \frac{\mu^2 M^2}{\sigma^4}\right)
   = \mu^4+\frac{\mu^4}{8} \left(\sigma^4 - \sigma_d^4
   \left(\frac{\sigma_d}{\sigma}\right)^4\right),
   \label{eq:effective_pot}
\end{equation}

\noindent
where $\sigma_d \equiv \sqrt{2}/3^{3/8}(\mu M)^{1/4}$.  If the scalar
potential is dominated by the first term, the inflaton $\sigma$ slow
rolls and therefore inflation occurs.

\bigskip

\noindent
According to the WMAP 5yr data, the curvature
perturbation ${\cal R}$, the spectral index $n_s$ and its running
$dn_s/d\ln k$ at the pivot scale $k_* = 0.002 {\rm Mpc}^{-1}$ are

\begin{eqnarray}
   {\cal R} & = & 4.9\times 10^{-5}, 
   	\label{eq:R_obs}\\
   n_s & = & 1.031\pm 0.055, 
   \label{eq:index_obs}\\
   \frac{dn_s}{d\ln k} & = &  -0.037\pm 0.028.
   	\label{eq:running_obs}
\end{eqnarray}
From the effective potential, we obtain
\begin{eqnarray}
   {\cal R} & = & \frac{V^{3/2}}{\sqrt{3}\pi V'} =\frac{\mu^2}{\sqrt{3}\pi}
   \left[ \sigma_*^3 + \sigma_d^3\left(\frac{\sigma_d}{\sigma_*}\right)^5
   \right]^{-1}, 
   \label{eq:R_hybrid}\\
   n_s -1 & \simeq  &  2\frac{V''}{V}= \left[ 
   3\sigma_*^2 - 5\sigma_d^2\left(\frac{\sigma_d}{\sigma_*}\right)^6
   \right], 
   \label{eq:index_hybrid}\\
   \frac{dn_s}{d\ln k} & \simeq  & -2\frac{V''' V'}{V^2}= -3\left[ 
   \sigma_*^3 + \sigma_d^3\left(\frac{\sigma_d}{\sigma_*}\right)^5
   \right]
   \left[ 
   \sigma_* + 5\sigma_d\left(\frac{\sigma_d}{\sigma_*}\right)^7
   \right],
   \label{eq:running_hybrid}
\end{eqnarray}
where $\sigma_*$ is the field value of the inflaton
when the fluctuation corresponding to the pivot scale exits the Hubble horizon.

\noindent
The fluctuation corresponding to the pivot scale $k_*$ exits the horizon at $t=t_*$ when 
$k_*/a(t_*) = H_{H} =\mu^2/\sqrt{3}$ ($H_{H}$: hubble during the smooth hybrid
inflation). Thus the scale factor $a_* = a(t_*)$ is given by
\begin{equation}
    \ln a_* \;=\; -2\ln \mu -136.
\end{equation}
The e-folding number between the horizon exit of the pivot scale
and the end of the smooth hybrid inflation is estimated as

\begin{eqnarray}
N_*(\sigma) & = & \int^{\sigma_*}_{\sigma_e} d\sigma \frac{V}{V'} \nonumber \\ 
& \simeq &  \frac{4}{3\sigma_d^2} -\frac{1}{\sigma_*^2}  ~~ (\sigma_* > \sigma_d)  \nonumber \\
& \simeq & \frac{\sigma_*^6}{3\sigma_d^8}  ~~ ( \sigma_* < \sigma_d) 
\label{eq:efold_hybrid}
\end{eqnarray}

\noindent
where $\sigma_e (\ll \sigma_d)$ denotes the field value when the
smooth hybrid inflation ends.

\bigskip

\noindent
After the smooth hybrid inflation, $\sigma$ and $\psi$ oscillate about
their minima and decay into the $\sigma$ and $\psi$ quanta via
self-couplings and mutual coupling of the two fields. Since their
effective masses depend on the field amplitudes and therefore
time-dependent, specific modes of the $\sigma$ and $\psi$ quanta are
strongly amplified by parametric resonance. To see this, let us write
down the evolution equation for the Fourier modes of fluctuations
$\sigma_k$ from (\ref{eq:pot_hybrid}) as
\begin{equation}
   \sigma_k'' + 3H \sigma_k' 
   +\left[\frac{k^2}{a^2} + m_{\sigma}^2 
   + 3m_{\sigma}^2\frac{\tilde{\psi}}{\sqrt{\mu M}}
   \cos (m_\sigma t)\right]\sigma_k\; \simeq \;0,
   \label{eq:mathieu}
\end{equation}

\noindent
where $m_{\sigma} = \sqrt{8\mu^3/M}$ and $\tilde{\psi}$ is the
amplitude of the $\psi$ oscillations. ($\tilde{\psi} \sim \sqrt{\mu
  M}$ at the beginning of the oscillations.)  Neglecting the cosmic
expansion, Eq.~(\ref{eq:mathieu}) has a form similar to the Mathieu
equation which is known to have a exponentially growing solution. The
detailed numerical simulation showed that the wave number for the
fastest growing mode is given by

\begin{equation}
  \frac{ k_p}{a_{\rm osc}}\; \simeq\; 0.3 \, m_\sigma.  
   \label{eq:k_peak}
\end{equation}

\noindent
The fluctuations amplified by the parametric resonance eventually
produce PBHs when they reenter the horizon after inflation. The mass
of the PBH is approximately given by the horizon mass when the
fluctuations reenter the horizon. Thus the PBH mass is estimated as

\begin{equation}
  M_{\rm BH}\; \simeq \; 1.4\times 10^{13} \, M_{\odot} \left(\frac{k_p}{{\rm Mpc}^{-1}}
  \right)^{-2}.
  \label{eq:BH_mass}
\end{equation}

\noindent
From Eqs.~(\ref{eq:k_peak}) and (\ref{eq:BH_mass}) the scale factor at
the beginning of the oscillation phase is estimated as

\begin{equation}
   \ln a_{\rm osc}\; =\; -114 -\ln m_\sigma -0.5 \ln (M_{\rm BH}/M_{\odot}).
\end{equation}

\noindent
Because the e-folding number $N_*$ is equal to $\ln a_{\rm osc} - \ln a_*$,  we obtain

\begin{equation}
   N_* \;=\; 21 + 0.5 \ln (\mu M) -0.5\ln(M_{\rm BH}/M_{\odot}).
   \label{eq:efold_hybrid2}
\end{equation}

\noindent
For a fixed black hole mass $M_{\rm BH}$, there are two parameters in
the model, i.e., $\mu$ and $M$, one of which can be removed by using
the WMAP normalization (\ref{eq:R_obs}).  Therefore observable
quantities can be expressed in terms of one free parameter, leading to
a non-trivial relation between $n_s$ and $d n_s/d \ln k$.  In
practice, we adopt $\mu M$ as the free parameter, and solve
Eqs.~(\ref{eq:efold_hybrid}) and (\ref{eq:efold_hybrid2}) for
$\sigma_*$ in terms of $\mu M$. Then $\mu$ and $M$ are determined with
use of Eqs.~(\ref{eq:R_hybrid}) and (\ref{eq:R_obs}) for a fixed $\mu
M$.  Thus, varying $\mu M$, we obtain sets of model parameters which
are consistent with the observed curvature perturbations.

\bigskip

\noindent
After $\sigma$ and $\psi$ decay, the second inflation ($=$ new
inflation) starts. As mentioned before, the role of the new inflation
is to stretch the fluctuations produced during the smooth hybrid
inflation and subsequent preheating phase to appropriate cosmological
scales. The effective potential for the new inflation is given by

\begin{equation}
  V_{\rm new} \;=\; v^4 \left( 1-\frac{c}{2}\phi^2 \right)-\frac{g}{2}v^2\phi^4
  +\frac{g^2}{16}\phi^8,
\end{equation}

\noindent
where $\phi$ is the inflaton of the new inflation, $v$ is the scale of
the new inflation and $g$ and $c$ are constants. The scale factor
$a_f$ at the end of the new inflation is estimated as
\begin{equation}
 \ln a_f  = -68 + \frac{1}{3} \ln \left( \frac{T_R}{10^9 GeV} \right)
 -\frac{4}{3} \left( \frac{v}{10^{15} GeV} \right),
\end{equation}
where $T_R$ is the reheating temperature after the new inflation. 
Therefore, the new inflation should provide the total e-fold number
$\simeq (\ln a_f - \ln a_{\rm osc})$.

\bigskip

\noindent
What makes the PBH particularly attractive as a DM candidate is that
it is naturally long-lived due to the gravitationally suppressed
evaporation rate. No discrete symmetries need to be introduced in an
ad hoc manner.  Also the PBH DM may be motivated from the arguments
based on entropy of the Universe~\cite{Frampton:2009nx}.

\section{The Solution for Dark Energy}

\noindent 
At the beginning of the twenty-first century,
there existed a problem in mathematics
which was considered so difficult that it was expected that
the century might end without solution. The problem was
the Poincar\'{e} Conjecture in topology.

\bigskip

\noindent 
In fundamental theoretical physics, there was,
at the beginning of the twenty-first century,
an equally impossible seeming problem which likewise
might not be solved for a hundred years. The problem
was the Dark Energy in cosmology.

\bigskip

\noindent 
The creativity of {\it homo sapiens} had been 
underestimated. The Poincar\'{e} Conjecture was proved
by Perelman, in less than three years. The Dark Energy
problem was solved, by myself, in less than ten years.

\bigskip

\noindent
In my Festschrift from 2003,
there is a photograph\cite{page19} of a four-year-old boy
with three special properties - a talent for mathematics,
infinite chutzpah and he said he is cleverer than Newton.
The talent
meant that if the young boy were given a three-digit
number, say, 506, he could, within seconds, answer 22x23; at most,
one per cent of four-year-olds could do, similarly.
At that time, in 1948, Newton was better known, even
than any of the monarchs, except possibly the then
monarch, King George VI. Surely, Newton was among the
top one percent of human intelligence, so to
be cleverer would require further reality checks. One
would be forthcoming in 1965.

\bigskip

\noindent
On the road from 1948 to 2010, I  will make mercifully brief rest stops
at 1957, 1965 and 2006. The first of these, 1957, is when I
learned, at King Charles I School, about the universal law
of gravitation. This was a key stage, because I clearly recall
looking up at the Moon and feeling my own weight, and being
so impressed by the idea that I decided, then and there, that
I would, one day, have a grander idea, than Newton's. At about
the same time, in 1957, my French teacher recommended, 
to my parents, a career, as a 
university professor, in linguistics. I might have done that,
were it not for the call of Newton.
Finally, in 1957, it was a memorable year because I met,
for twelve seconds in Kidderminster Town Hall,
the monarch, Queen Elizabeth II. Having
bowed, I was ready to answer absolutely any question but
all she said was that it was very nice to meet me. I should
have worn a sign, soliciting a royal question.

\bigskip

\noindent
In 1965, it was my turn for the opportunity of the Oxford Final
Honors Schools (OFHS) with its six three-hour examinations, two
each on Wednesday, Thursday and Friday June 6 - 8,
1965. The three morning exams were conventional while
the afternoon OFHS exams were open-ended essay questions,
with no instruction, even on how many questions to answer.

\bigskip

\noindent
For the four months February - May, 1965 I did nothing, except
study and make extensive notes, and memory cards. I was 
sequestered, in Frewin Hall, and talked to 
nobody, except college servants who could bring me food, or
physics books from Blackwell's.
What is pertinent to the sequel, in 2007 and 2010,
is that of the hundred physics books I accumulated in Frewin Hall, my personal
favorite was always Tolman's {\it Relativity, Thermodynamics and Cosmology},
a clear and endearingly modest discussion, of the role of
entropy in cyclic cosmology. I do recall spending hours then intrigued by
the apparent contradiction, between the attractive idea
of cyclic cosmology, and the second law of themodynamics; the
contents of Tolman's book, however, did not appear on my
examinations.

\bigskip

\noindent
For the OFHS paper on Thursday afternoon (June 7, 1965) my strategy
was to answer only one essay question. I had retained extensive material
on a dozen topics, with a good probability at least one of them would appear
on the question paper. There it was: X-ray diffraction. In three hours,
I produced a meticulously-detailed 100-page monograph on X-ray diffraction,
later described by an experienced examiner, as the most detailed answer,
he had ever seen.
This required some of Gell-Mann's attributes: clear thinking, profound understanding
and extensive retention. Incidentally, it also needed fast handwriting. My OFHS grades
on my six papers were $\alpha, \alpha, \alpha, \alpha^+,\alpha, \alpha$.
This is called straight alphas. Two alphas were necessary for First Class Honours. 
The unprecedented $\alpha^+$ led to some discussion, in the
Brasenose College (BNC) senior common room, and the BNC
Fellows decided to allow me dining rights, on High Table, for as long as
I would remain at BNC, as a doctoral student. The $\alpha^+$ did support 
my being in the top one percent of human intelligence, just like Newton.
At High Table dinners,
 I befriended a philologist ,who had collected numerous honorary doctorates, and
 could understand a hundred languages. He once mentioned that he had met,
 dining in BNC, just the previous evening, Gell-Mann who had explained his ideas,
 about the origin, of the Basque language. Therefore, I could have first met Gell-Mann
 in 1965, in BNC, had I attended that dinner. Instead I first met Gell-Mann
 the following year, 1966, as discussed in the next section.
 
 \bigskip
 
 \noindent
 More than fourty years after my OFHS experience, and after the accelerated
 cosmic expansion had been discovered, in 1998, I took on
 a new PhD student at UNC-Chapel Hill,  Lauris Baum, in 2006 and suggested
 that he study, assiduously, existing papers on cyclic cosmology. This he did,
 and we discussed, at length, the issue of the Tolman conundrum, which
 had first piqued my intellectual curiosity, in 1965. The result was the first, and still
 only, solution to the 75-year-old conundrum\cite{Tolman,BF}.
 In 2010, at Tokyo, on Thursday, February 4, Hirosi Ooguri who is a distinguished professor
 at the California Institute of Technology and, like me, a professor at
 the University of Tokyo (I am also a distinguished professor in Chapel Hill)
 wrote, to inform me
 \footnote{A useful communication, from Ooguri san, at IPMU, is acknowledged.}, that, on Saturday,
 a Todai visitor, Professor Dam Son, would give three lectures
 on the holographic principle at Hongo campus, starting at 1:30 PM. 
 Son's lectures exceeded expectations.
 During the lectures (February 6, 2010), I realized, writing in my notebook, that
 the visible universe is approximated by a black hole, and that this leads to
 a resolution of the dark energy problem\cite{Frampton}.

\bigskip

\noindent
Consider the Schwarzschild
radius ($r_s$), and the physical radius ($R$), of the Sun ($\odot$).
They are $(r_s)_{\odot} = 3 km$ and $R_{\odot} = 800,000 km$.
Their ratio is $(\rho)_{\odot} \equiv (R/r_s)_{\odot} = 2.7 \times 10^5$.
One can readily check that, for the Earth or for the Milky Way, that
the ratio $\rho = (R/r_s)$ is likewise much larger than one: $\rho >> 1$.
Such objects are nowhere close to being  black hole. Now 
consider the visible universe (VU), with mass $M_{VU} = 10^{23}
M_{\odot}$. It has $(r_s)_{VU} = 30 Gly$, and $(R)_{VU} = 48 Gly$,
hence $(\rho)_{VU} = 1.6$. The visible universe, within which we all live,
is close to being a black hole. The solution to the dark energy
problem follows, providing I so approximate the
visible universe. At the horizon, there is a PBH
temperature \cite{Parker, Bekenstein,Hawking}, $T_{\beta} $, which I can estimate as

\begin{equation}
T_{\beta} = \frac{\hbar}{  k_B }~\frac{H }{ 2 \pi} \sim 3 \times 10^{-30} K .
\label{T-beta}
\end{equation}

\bigskip

\noindent This temperature of the horizon information screen leads
to a concomitant  FDU acceleration \cite{Fulling,Davies,Unruh}
 $a_{Horizon}$, outward, of the horizon given
by the relationship

\bigskip

\begin{equation}
a_{Horizon} = \left( \frac{2 \pi c k_B T_{\beta}}{\hbar} \right) = c H \sim 10^{-9} \, m/s^2
\,.
\label{acceleration}
\end{equation}

\bigskip

\noindent When $T_{\beta}$ is used in Eq. (\ref{acceleration}),
I arrive at a cosmic acceleration which is essentially in
agreement with the observations\cite{Perlmutter, Reiss}.

\bigskip

\noindent
It would be a wonderful to have lunch, may be at L'Atelier de
Jo\"{e}l Robuchon 
in Roppongi Hills, with Murray Gell-Mann, Isaac Newton,
and Grigori Perelman to compare notes on personal
fulfillment. What does Grigori Perelman mean, when he tells journalist, in
turning down a million dollars, {\it I have all I want. I'm not interested in money or fame}?
This seems to baffle some americans, whose idea of happiness, as an
inalienable right, is  a three-comma net worth.
Yet, a two-comma net worth suffices, for all practical purposes.
Fame can hardly exceed that of the singer and entertainer, Elvis Presley (1935-1977), whose name, from my non-scientific
studies in public transportation, is still recognizable by
one billion people.
He died, when he was only fourty-two, so his fame was not very useful.

\bigskip

\noindent
After Son's lectures on February 6, 2010, I went to the nearby 
Yushima Shrine around 6:00 PM and, impossibly, hoped that one of the
many Japanese strolling around the shrine was Nambu sensei, to tell him.
One ramification was that most of the work on quantum gravity, since the
discovery of quantum mechanics, was called into question.
There was an indescribable feeling of personal fulfillment, that the 66
years and 98 days, so far, of my life, had a significance.
This was/is a totally individual experience which, unlike money or fame,
involves no other person, and is therefore different. Because the visible 
universe is much bigger than the Solar System
\footnote{A useful discussion, with Gerard 't Hooft, at the Gell-Mann Festschrift, is acknowledged.}, 
I had vindicated my claim, as a four-year-old, to be cleverer than Newton.
Because, in my opinion, time travel into the past will forever be impossible,
I cannot return to Isaac Newton in 1686 and forewarn him that a cleverer
person will be born on October 31, 1943; nor can I return to 1948 and tell
the four-year-old on a tricycle that he is right to say
he is cleverer than Newton.
The first reaction is to want to achieve the personal fulfillment again, and again.
I am certain that Perelman is presently pursuing the six other Clay prolems,
in alphabetical order: Birch and Swinnerton-Dyer Conjecture, Hodge Conjecture,
Navier-Stokes Equations,
P vs NP, Riemann Hypothesis and Yang-Mills Theory. More
likely, Perelman is considering a more profound direction,
in mathematics.

\bigskip

\noindent
Newton finished Book I of {\it Principia}, entitled
{\it De Motu Corporum I}, in 1686; then Book II
({\it De Motu Corporum II}) and Book III ({\it De Systemata Mundi}) in 1687.
Book III adds more empiricism. I now explain
why PHF would write, even then in 1685, a better {\it Principia}
than Newton. PHF would start, in 1685, with Book II (Newton's grade, B-),
knowing that Book I (Newton's grade, A+) was easier.

\bigskip

\noindent
In order to explain why his sound speed formula $v_s = \sqrt{p/\rho}$
gives $v_s = 290 m/s$ whereas the experimental value
for $v_s$ at one atmospheric
pressure and $T=20^0 C$ is $v_s = 343 m/s$, Newton needed a large
correction. About a half of this correction arises, according to {\it Principia} Book II,
from Newton's crassitude, where the sound
propagates instantaneously through 
particles in the air. The remaining discrepancy
leads to, surely, the most confused passage, in all of the
{\it Principia}. Although I know that Gell-Mann
\footnote{A useful discussion, with Murray Gell-Mann, at the Gell-Mann Festschrift, is acknowledged.}
 reads Latin as well as I do, other non-British-educated
theoreticians may not, so I quote, instead of Latin, an English translation of
a Scholium:

\bigskip

\noindent
{\it Moreover, the vapors floating in the air, being of another
spring, and a different tone, will hardly, if at all, partake of
the motion of the true air in which the sounds are propagated. Now
if these vapors remain unmoved, that motion will be propagated
the swifter through the true air alone,
and that in the subduplicate
ratio of the defect of the matter. So if the atmosphere
consists of ten parts of true air
and one part of vapors,
the motion of sounds will be swifter
in the subduplicate ratio of 11 to 10, or very nearly
in the entire ratio of
21 to 20, than if it were propagated through eleven parts of true air: and therefore the motion
of sounds above discovered must be increased in that ratio.}

\bigskip

\noindent
Newton was not only clever in mathematics, he was
also a brilliant experimentalist. He himself measured the speed of sound
in Nevile's Court
\footnote{A useful discussion, with Bernard Carr, at IPMU, is acknowledged.}
at Trinity College in Cambridge by hitting
the paving stone with a hammer at such a frequency 
that the echo coincided with the next hit. Other experimentalists,
such as Sauveur, cited by Newton,
had determined $v_s$, so there was no doubt the 
theory was wrong.

\bigskip

\noindent
The cleverer PHF would have 
thought more deeply, than Newton, about the ratio
$r = (v_s)_{expt}/(v_s)_{theory}$.
It is not too difficult to see, that this requires the isothermal
Boyle's equation of state for an ideal gas, $PV = constant$ to
become the adiabatic $PV^{r^2} = constant$.
From theory and experiment, one knows $r^2 = 7/5$
and then, via diatomic molecules and statistical mechanics,
I arrive smoothly at the entropy
defined by Clausius, whose birthname
\footnote{A useful discussion, with Finn Ravndal, at the 
Gell-Mann Festschrift, is acknowledged.} was Gottlieb,
in 1865. What emerges is Boltzmann's equation $S=k \ln W$
as more profound than the equations of Newton,
like $F=Gm_1m_2/R^2$, or those
of the, still future, Einstein, like $E=m c^2$. Here the emphasis is not
on exactitude, but on profundity
\footnote{A useful discussion, with Murayama san, at
IPMU,
is acknowledged.} .

\bigskip

\noindent
The aforementioned solution, of the dark energy problem, not only
solves a cosmological problem, it casts a completely new light, on the
nature of the gravitational force. Since the expansion of the universe,
including the acceleration thereof,
can only be a gravitational phenomenon, I arrive at the viewpoint, that
gravity is a classical result, of the second law of thermodynamics.
This means that gravity cannot be regarded as, on a footing
with, the electroweak and strong interactions.
Although this can be the most radical change, in gravity
theory, for over three centuries, it is worth emphasizing,
that general relativity  remains unscathed.

\bigskip

\noindent
My result calls into question, almost all of the work done
on quantum gravity, since the discovery
of quantum mechanics. For gravity, there is no longer necessity
for a graviton. In the case of string theory, the principal
motivation\cite{yoneya,ScherkSchwarz} for the 
profound, and historical, suggestion,
by Scherk and Schwarz, that string theory be reinterpreted,
not as a theory of the strong interaction, but instead as
a theory of the gravitational interaction,
came from the natural appearance, of a massless graviton,
in the closed string sector.
I am not saying that string theory is dead.  What I am
saying is, that string theory cannot be a theory of the
fundamental gravitational interaction, since there is
no fundamental gravitational interaction.

\bigskip

\noindent
The way this new insight emerged, and the solution of
the dark energy problem itself, was as a natural line
of thought, following the discovery of a cyclic model in
\cite{BF}, and the subsequent investigations
\cite{Frampton:2009nx,Entropy1,Entropy2,EFS1,EFS2} of the
entropy
of the universe, including a possible candidate
for dark matter\cite{Frampton:2009nx,FKTY}.

\bigskip
\noindent
Another ramification, of my solution of the dark energy,
problem is the status, fundamental versus emergent,
of the three spatial dimensions,
that we all observe every day. Because the solution
assumes the holographic principle\cite{Hooft},
at least one spatial dimension appears as emergent
\footnote{A useful discussion, with John Schwarz, at
the Gell-Mann Festschrift, is acknowledged.}
\footnote{A useful discussion, with Sugimoto san, at IPMU, is acknowledged.}.
Regarding the visible universe as a sphere,
with radius of about 48 Gly, the emergent space
dimension is then, in spherical polar coordinates,
the radial coordinate, while the other two coordinates,
the polar and azimuthal angles, remain fundamental.
Physical intuition, related to the isotropy of space,
may suggest that, if one space dimension is emergent,
then so must be all three. This merits further investigation,
and may require a generalization of the
holographic principle in \cite{Hooft}.
On the other hand, a fundamental time
coordinate is useful in dynamics.
This present discussion
is merely one step towards the goal of a
cyclic model, in which time never begins or ends.

\section{Gell-Mann in Twentieth Century Physics}

\noindent 
Whereas I have published research in particle phenomenology for fourty
years, and whereas I will not include my own name in the list,
these are sufficient credentials to assess the greatest theoreticians
of the twentieth century. I shall arrive, at a top-ten list, which includes: four
very distinguished Europeans, four truly brilliant Americans all born, by coincidence, in the great state of New York and
two living-legend Asians, of whom, only Yang has been significantly influenced, in adult life, by
the Confucian analects
\footnote{A useful discussion, with Yang Zhenning, 
at the Gell-Mann Festschrift, is acknowledged.}.

\bigskip

\noindent
In alphabetical order, the top five, with two chosen accomplishments:

\bigskip

Paul Dirac ({\tt antimatter} \underline{\it and} {\tt g = 2})

Albert Einstein ({\tt relativity} \underline{\it and} {\tt photoelectric effect})

Murray Gell-Mann ({\tt $\Omega^{-}$} \underline{\it and} {\tt quarks})

Gerard 't Hooft ({\tt holographic principle} \underline{\it and} {\tt renormalizability})

Yang Zhenning ({\tt gauge theories} \underline{\it and} {\tt parity violation})

\bigskip

\noindent
The next five are, again in alphabetical order: Richard Feynman, Sheldon Glashow, 
Werner Heisenberg, Nambu sensei and Julian Schwinger.
Below these, the ordering becomes more subjective, but my top ten
choices, I believe, are close to the general opinion.

\bigskip

\noindent
It should be noted that, in 1948, Nambu sensei, independently 
of the late Julian Schwinger, derived the 
one-loop quantum electrodynamics correction to
$(g-2)$. That would give Nambu sensei ({\tt symmetry breaking}
\underline{\it and} {\tt (g-2)})
which is very strong, and could displace one of the top five. However, Nambu sensei did not
\footnote{A useful discussion, with Nambu sensei, at Osaka University, is acknowledged.} publish,
possibly because he did not want to overshadow Tomonaga sensei, fourteen years senior, chronological age 
being all-important in Japanese society.

\bigskip

\noindent
I first met Murray (if I may) in 1966 when I was starting research
in particle phenomenology and my Oxford doctoral adviser, J.C. Taylor, considered
it worth driving ten miles to the Rutherford Laboratory. It was indeed
worthwhile. Murray spoke with infinite self-confidence, and, in answering 
questions, provided information, like a computer download, reflecting
encyclopaedic knowledge. In
those times, Murray's prescient paper \cite{MGM1962}  {\it Symmetries of Baryons and
Mesons} was a standard reference for Oxford students.

\bigskip

\noindent
Murray has many first-rate accomplishments. Equally impressive, is
the sheer number of new results, sometimes several in the same year
\footnote{ A useful discussion, with Kenneth Wilson, at the Gell-Mann Festschrift, is acknowledged.}
of which I can mention, in the time available, just a hint
of Murray's gigantic contributions, with the renormalization group\cite{Low},
the sigma model\cite{Sigma} and the invention\cite{QCD} of the
theory of strong interactions.  As one speaker at this Festschrift put it,
everything in particle phenomenology was either by
Murray, or named by Murray, who enriched the field, with
erudite names, like strangeness, and quark\cite{quarks}. In his
monumental quark paper, perhaps the best two pages ever
printed in Physics Letters B, Murray's infinite self confidence
wobbled, when he discussed {\it non}-existence of real quarks.

\bigskip

\noindent
Murray, I wish you many more years
of creativity. You are, forever, a giant in particle phenomenology.

\bigskip

\noindent
\section{Acknowledgements}

\noindent
This work was supported in part by
the World Premier International Research Center Initiative 
(WPI initiative), MEXT, Japan and by U.S. 
Department of Energy Grant No. DE-FG02-05ER41418.

\end{document}